\documentclass[galaxies,article,accept,moreauthors,pdftex,10pt,a4paper]{mdpi}
\pdfoutput=1

\firstpage{1}
\articlenumber{8}
\doinum{10.3390/galaxies5010008}
\pubvolume{5}
\pubyear{2017}
\copyrightyear{2017}
\externaleditor{Academic Editors: Jose L. G\'{o}mez, Alan P. Marscher and Svetlana G. Jorstad}
\history{Received: 13 September 2016; Accepted: 16 January 2017; Published: date}

\usepackage{soul,microtype}

\Title{Proper Motions of Jets on the Kiloparsec Scale: New~Results with HST}

\Author{{Eileen T. Meyer}~$^{1,}$*, William B. Sparks~$^{2}$, Markos Georganopoulos~$^{1}$, Roeland van der Marel~$^{2}$, Jay~Anderson~$^{2}$, Sangmo T. Sohn~$^{2}$, John Biretta~$^{2}$, Colin Norman~$^{2}$, Marco Chiaberge~$^{2}$ \mbox{and Eric Perlman~$^{3}$} }

\AuthorNames{Eileen Meyer, William Sparks, Markos Georganopoulos, Roeland van der Marle, Jay Anderson, Sangmo Sohn, John Biretta, Colin Norman, Marco Chiaberge, Eric Perlman}

\address{%
$^{1}$ \quad Department of Physics, University of Maryland, Baltimore County, Baltimore, MD 20742, USA; georgano@umbc.edu\\
$^{2}$ \quad Space Telescope Science Institute, Baltimore, MD 21218, USA; sparks@stsci.edu (W.B.S.); marel@stsci.edu (R.v.d.M.);  jayander@stsci.edu (J.A.); tsohn@stsci.edu (S.T.S); john.biretta@gmail.com (J.B.); norman@stsci.edu (C.N.); marcoc@stsci.edu (M.C.) \\
$^{3}$ \quad Florida Institute of Technology, Melbourne, FL 32901, USA; eperlman@fit.edu\\
}
\corres{Correspondence: meyer@umbc.edu}

\abstract{The Hubble Space Telescope recently celebrated 25 years of operation. Some of the first images of extragalactic optical jets were taken by HST in the mid-1990s; with time baselines on the order of 20 years and state-of-the-art astrometry techniques, we are now able to reach accuracies in proper-motion measurements on the order of a tenth of a milliarcsecond per year.  We present the results of a recent HST program to measure the kiloparsec-scale proper motions of eleven nearby optical jets with Hubble, the first sample of its kind.  When paired with VLBI proper-motion measurements on the parsec scale, we are now able to map the full velocity profile of these jets from near the black hole to the final deceleration as they extend out into and beyond the host galaxy.  We~see convincing evidence that weak-flavor jets (i.e., FR Is) have a slowly increasing jet speed up to 100~pc from the core, where superluminal components are first seen.}


\begin{document}



\section{Introduction}

Many aspects of the physics of relativistic jets from super-massive black holes are not well understood, including the particle makeup of the jet, their lifetimes, and the speed profile of the plasma as it extends out of the host galaxy and into the intergalactic medium.
Proper-motion studies allow us to directly observe the apparent speeds
of these jets, resulting in constraints on their Lorentz factors
($\Gamma$). Hundreds of observations of jets with very long baseline
interferometry (VLBI) in the radio have detected proper motions of
jets on parsec and sub-parsec scales, relatively near to the black
hole engine
(e.g.,~\cite{giovannini2001,piner2004,kellermann2004,jorstad2005,lister2009}). Because these jets are highly relativistic, apparent superluminal motions can result from velocities near the speed of light coupled with relatively small viewing angles. We start with some definitions: the
dimensionless observed apparent velocity $\beta_\mathrm{app}$ is
related to the real velocity $\beta=v/c$ (where $c$ is the speed of
light) and viewing angle $\theta$ through the well-known Doppler
formula $\beta_\mathrm{app}=\beta \sin \theta/(1 - \beta \cos
\theta)$. A measurement of $\beta_\mathrm{app}$ implies both a lower
limit on the Lorentz factor
($\Gamma_\mathrm{min}\approx\beta_\mathrm{app}$) and an upper limit on
the viewing angle. These constraints are difficult to derive
using other means such as spectral fitting, due to the inherent
degeneracy between intrinsic power, angle, and speed introduced by
Doppler boosting of the observed flux.

While there are large samples of proper-motion measurements for jets on parsec scales,
direct observations of jet motions on much larger scales (kpc--Mpc) with high-resolution
telescopes such as the Very Large Array (VLA) or Hubble Space Telescope (HST) are rare. However, the much lower-than-VLBI resolution necessarily also limits potential observations of
apparent motions to sources in the very local Universe, and often require
years or even decades of repeated observations. For many years, there were only two measured proper motions of jets on
kpc scales, both taken with the VLA. These were the famous result of
$\beta_\mathrm{app}$ up to $6c$ measured by \cite{biretta1995} for the
jet in M87 \mbox{(z~=~0.004, d~=~22 Mpc),} and a speed of $\approx$4$c$ for a knot
in the jet in 3C 120 (z~=~0.033, d~=~130 Mpc) by~\cite{walker1988}, though
this was later contradicted by additional VLA and Merlin observations
\citep{muxlow1991,walker1997}. In 1999, the first measurement of
proper motions in the optical was accomplished by \cite{biretta1999},
using four years of HST Faint Object Camera (FOC) imaging to confirm
the fast superluminal speeds in the inner jet of M87.  In this paper, we present some recent proper-motions results for three nearby optical jets obtained from archival and new HST data, enabled in major part by the long operating lifetime of HST and the resulting 20-year baselines from the original snapshot images of optical jets first taken in the 1990s.

\section{Methods}

All three optical jets were observed with several of the Hubble Space Telescope imaging instruments over a span of 10--20 years.  The analysis of the new and archival data was similar for all three jets.  We will present first the methods for the jet in M87, and then briefly remark on the data used for 3C~264 and 3C~273, which was treated very similarly.  The data used for all jets and further details can be found in \cite{mey13_m87,mey15_nature,mey16_3c273}.

\subsection{M87}
\unskip
\subsubsection{Astrometry}
For M87, the reference frame is based on hundreds of bright globular
clusters associated with the host galaxy which are effectively
stationary to proper motions over the time of this study. The~reference frame was built using the Advanced Camera for Surveys Wide Field Camera (ACS/WFC) exposures in the F606W
filter, by first detecting the
positions of the globular clusters in each flat-fielded, CTE-corrected
ACS/WFC image using a PSF-peak-fitting routine, then applying the
standard (filter-specific) geometric correction to those positions,
and then finding the best linear transformation for each image which
matches the positions in the individual (geometrically-corrected)
frames to a master reference frame. This last step is done by a
routine similar to MultiDrizzle, but~better optimized for astrometry
(see~\cite{and10,sohn12}). The process of finding transformations
for all frames is iterated so that the master reference frame
(super-sampled to a pixel scale of 0.025$''$/pixel) effectively
gives the average position of each reference source in a
geometrically-corrected frame.

The final reference system consists of positions and magnitudes for
over 1300 globular clusters within $\sim$100$''$ of the M87 core
position. Among all clusters, the median one-dimensional RMS residual
relative to the mean position was 0.05 (reference-frame) pixels (1.25
mas), corresponding to a systematic astrometric accuracy ($\times$
1/$\sqrt 56$) of 0.17 mas; over the 13.25 year baseline this is
equivalent to 0.003$c$. Astrometric solutions were then found for all the ACS High Resolution Camera (ACS/HRC),  ACS/WFC,
and Wide Field Planetary Camera 2 (WFPC2) images in the F814W filter. Typical numbers of globular
clusters used to match the frames were $\sim$200--500 for ACS/WFC,
$\sim$15--30~for WFPC2 images, and $\sim$15--25 for ACS/HRC;
corresponding systematic errors were $\sim$0.006, $\sim$0.03, and
$\sim$0.05~pixels, or~0.003$c$, 0.015$c$, 0.025$c$ over 13.25 years,
respectively.

\subsubsection{Measuring the Knot Positions}
Within the M87 jet, we first identified the significant features (peaks) of interest using stacked
images. The image stacks include the 2006 ACS/WFC stack, four WFPC2
stacks (epochs 1995--1996, 1998--1999, 2001, and 2007--2008), and three
stacks of ACS/HRC exposures (November~2002--December~2003, July 2004--September 2005, November 2005--November 2006). The~host galaxy was
modeled using the ACS/WFC stacked image with the IRAF/STSDAS tasks
\emph{ellipse} and \emph{bmodel} and then subtracted.
A~two-dimensional continuous functional representation of the image
cutouts was then created using the Cosine Transform function
(\emph{FourierDCT}) in \emph{Mathematica}, which allowed us to find
prominent peaks, as well as contours of constant flux level around
those peaks.  In general, the number of interesting peaks was chosen
by hand, and the contour line levels were at 50\% of the flux of the
peak, after a ``background'' level was subtracted. The positions defining the contour were
reverse-transformed from master frame coordinates into each
distorted, galaxy-subtracted image so that an intensity-weighted
centroid position from the pixels within the contour could be
calculated.  These final positions were then transformed back into the
common reference frame, so that each peak of interest was measured in
every exposure, resulting in over 400 position measurements spread
over the 13 year baseline.

To measure the apparent motion of the jet components over time, we
define the x-direction as the line from the core through the center of
knot I (PA of 290$^\circ$), with y-direction perpendicular. Data~were
binned into 26 time bins (shown on the right in {Figure}~2), and a
linear model was fit using standard weighted least squares.  The
weights were taken to be the inverse of the variance for each time
bin, as~measured from 10 nearby globular cluster reference sources.

\subsection{3C~264}

We obtained deep V-band imaging of 3C 264 with the ACS in May of 2014 in order to compare nearly 20 years of
images taken by HST for evidence of proper motions of these knots. In a very similar approach as with M87, we
localized over 100 globular clusters in the host galaxy as a reference
system on which to register previous images taken with WFPC2 in 1994, 1996, and 2002. The systematic
error in the registration of the WFPC2 images is generally on the
order of 5 milliarcseconds (mas) or less.

To measure their apparent speeds, the position of each knot was
measured with a centroiding technique. In the case of knots B and C,
we also modelled the jet as a constant-density conical jet with
superimposed resolved knots, in order to better measure their fluxes
and positions, particularly in the final epoch when they appear to
overlap. We plotted the position of
each knot along the direction of the jet axis versus time, and fitted
the data with a least-squares linear model. A slope significantly
larger than zero indicates significant proper motions, and we used the
conversion factor 1.442$c$ year/mas to convert angular speeds
($\mu_{app}$) to units of $c$.

\subsection{3C~273}

We obtained four orbits of ACS/WFC imaging in F606W in May of 2014. These were
stacked into a mean reference image (with cosmic-ray rejection) on a
super-sampled scale with 0.025$''$ pixels. The registration of the eight
individual exposures utilized a full 6-parameter linear transformation
based on the distortion-corrected positions of 15--20 point-like
sources. The median one-dimensional rms residual relative to the mean
position was 0.07 reference-frame pixels, or 1.75 mas, corresponding
to a systematic error on the registration ($\times 1/\sqrt 16$) of
0.44 mas, or about two hundredths of a pixel.

\subsubsection{Background Source Registration}

\begin{figure}[H]
\centering
\includegraphics[width=8cm]{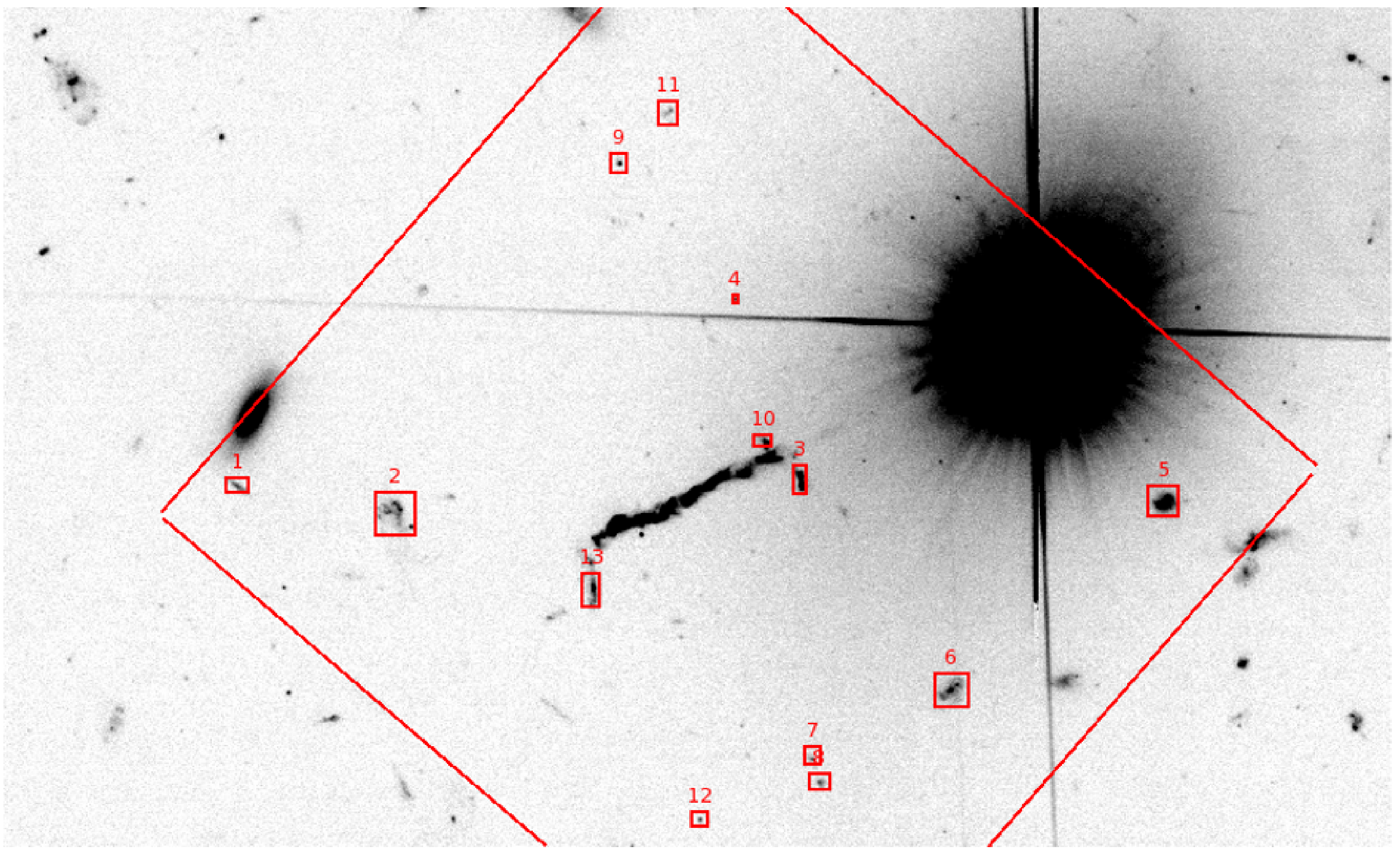}
\caption{The immediate environment of 3C~273 as seen in the ACS/WFC 
	 reference image from 2014. The background galaxies used for image registration are boxed. The larger outline is approximately the image footprint of the earlier 1995 WFPC2 
	 imaging. (Note that the image is in the native CCD 
	 orientation, and North is approximately to the right.)}\label{f1}
\end{figure}

Unlike with M87 and 3C~264, the globular clusters of 3C273 are not detectable in the older WFPC2 imaging and thus cannot be used as a reference frame. Instead, we identified 15 background galaxies
based on the criteria that they can be seen by eye above the noise in
the individual WFPC2 exposures.  These reference galaxies are highlighted
in Figure~\ref{f1}. Note that galaxies 4, 5, and 6
have been previously identified as unrelated to the jet by their lack
of radio emission, and the bright point source near the jet is
actually a foreground star (and thus unsuitable for registering
images due to likely proper motions).

To match the archival images to a common reference frame, our general
strategy was to use the shape and light distribution of the galaxy to
assist in matching their locations in each image. Instead of
identifying a single location associated with each galaxy in the deep
reference image, we~instead sample the galaxy in a grid pattern,
resulting in a list of positions along with the flux at each point,
sampling across each galaxy. An
initial astrometric transformation solution was found by supplying
$\approx$10 pairs of matched locations found by hand between the geometrically-corrected raw
image and the ACS reference image and calculating the six
transformation parameters (without~match evaluation/rejection).
The initial transformation solution described above is used as a
starting point to transform the x,y locations for each galaxy grid in
the reference frame into $x_{gc},y_{gc}$ location in the geometrically
corrected image. The intensity can then be sampled at each
location in the raw image, to~be compared directly to the scaled counts
value predicted by the scaled reference pixel value. This~information is used to calculate the optimal position shift which is used to update the location of the
galaxy in the reference frame.

For each individual exposure, we then compile an updated list of
position matches between the reference frame and the GC image from the
mean x,y value of the galaxy sampling grid in each. In~general, we used a subset
of the background galaxies which were identifiable by eye and not
overly affected by cosmic ray hits. The process of finding the initial
$x_{gc}, y_{gc}$ values, followed by finding the optimal
$\delta$x,$\delta$y improvement on the mean position, was iterated
until the positions of the galaxies stopped improving.

\subsubsection{Measuring Knot Positions in 3C~273}
As a consistency check, we also measured the shift of each knot using
a second method which we refer to as the ``cross-correlation''
method. Over a grid with sub-pixel spacing of 0.2 super-sampled pixels
(5 mas), we shifted the 1995 and 2003 images of each individual knot
relative to the 2014 image (same cutout area) over a 6~$\times$~6 pixel area,
evaluating the sum of the squared differences between interpolated
flux over the knot area for each x/y shift combination. The resulting
sum-of-squared differences image in all cases clearly shows a smooth
``depression'' feature which is reasonably well-fit by a two-dimensional
Gaussian under the transformation $g = 1 - f/\mathrm{max}(f)$, where
$f$ is the original sum of squared differences. Taking the minimum $f$
location as measured by the peak of the two-dimensional Gaussian fit,
we measure the optimal shift for each knot.

To measure the approximate error on the positions measured, we
repeated both of the above methods for simulated images of the jet at
each epoch. The simulated images were created by taking the deep 2014
ACS image and adding a Gaussian noise component appropriately scaled
from the counts in the original WFPC2 exposures. Since the 2014 image
itself has some noise, and also a slightly different PSF from the
WFPC2 images, this method likely slightly overestimates the errors. We
take the error on each knot measurement to be the standard deviation
of the measurements in the simulated images (10 in each epoch).

Finally, we plotted the position of each feature relative to the 2014
position, versus time, to look for evidence of proper motions. We have
transformed from the coordinate frame of the aligned images (North up)
to one based on the jet, where positive $x$ is in the outflow
direction along the jet (taken as position angle (PA) 42$^\circ$ south of
east) and positive $y$ is orthogonal and to the north of the jet. For both methods, the estimated
error on the measurement has been convolved with the systematic error
of the registration, which is 0.18, 0.22, and 0.02 reference pixels
(4.5, 2.8, and 0.5 mas) for the 1995, 2003, and~2014 epochs,
respectively.

\section{Results and Discussion}
\unskip
\subsection{M87}

\begin{figure}[H]
\centering
\includegraphics[width=13cm]{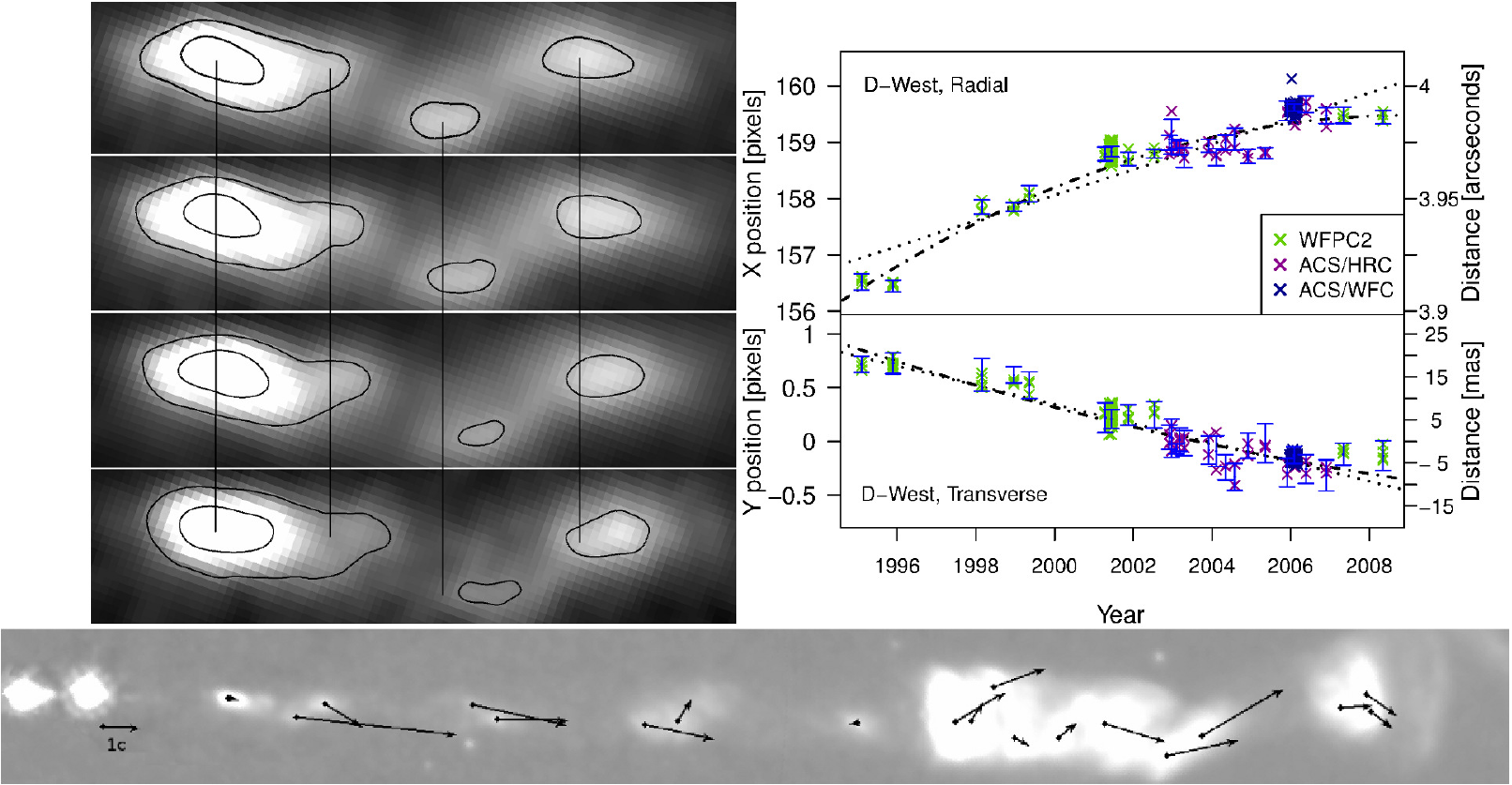}
\caption{Some results from the M87 study.  At upper left, a close view of knot D complex over four epochs from 1995 to 2008.  As shown, the Eastern component (far left) extends but does not change the core location, while the two sub-knots to the right show both radial motion and deceleration. On the right, the position versus time of the Westernmost component of knot D. In the lower panel, we show the superimposed velocity vectors on each component in the M87 jet. Length corresponds to the magnitude of the total speed.}\label{f2}
\end{figure}

The most notable findings for M87 were the findings of both deceleration and transverse motions for the first time.  In Figure~\ref{f2}, we show the knot complex D, which consists of three sub-knots.  Knot D is the most consistently bright feature after the knot A/B/C
complex, and significant proper motions were measured in previous
studies. As suggested by the contours overlaid on Figure~\ref{f2},
D-Middle is one of the fastest components with a speed of
4.27~$\pm$~0.30$c$ along the jet, while the brighter D-West shows
evidence of \emph{deceleration} radially, slowing to a near stop by
the final epoch in 2008, while maintaining one of the largest
transverse speeds of $-$0.59~$\pm$~0.05$c$ (right panel of Figure~\ref{f2}).
Previous results on knot D-East have been conflicting: B95 found that
it moved \emph{inward} along the jet at 0.23~$\pm$~0.12$c$ (possibly~consistent with being stationary), while B99 found a large outward
apparent velocity of 3.12~$\pm$~0.29$c$; our result of 0.28~$\pm$~0.05 is
more in line with B95. It is possible that there is a stationary
feature at D-East, through which components emerge (analogous to what
is seen in HST-1). In that case, the higher-resolution FOC was perhaps
tracking an emerging bright component, while over longer periods the
global feature at D-East is stationary.  In the outer jet (from knot A), we~find apparent velocities that are still superluminal and velocity vectors that appear to line up into a helical/side-to-side pattern.

\begin{figure}[H]
\centering
\includegraphics[width=15cm]{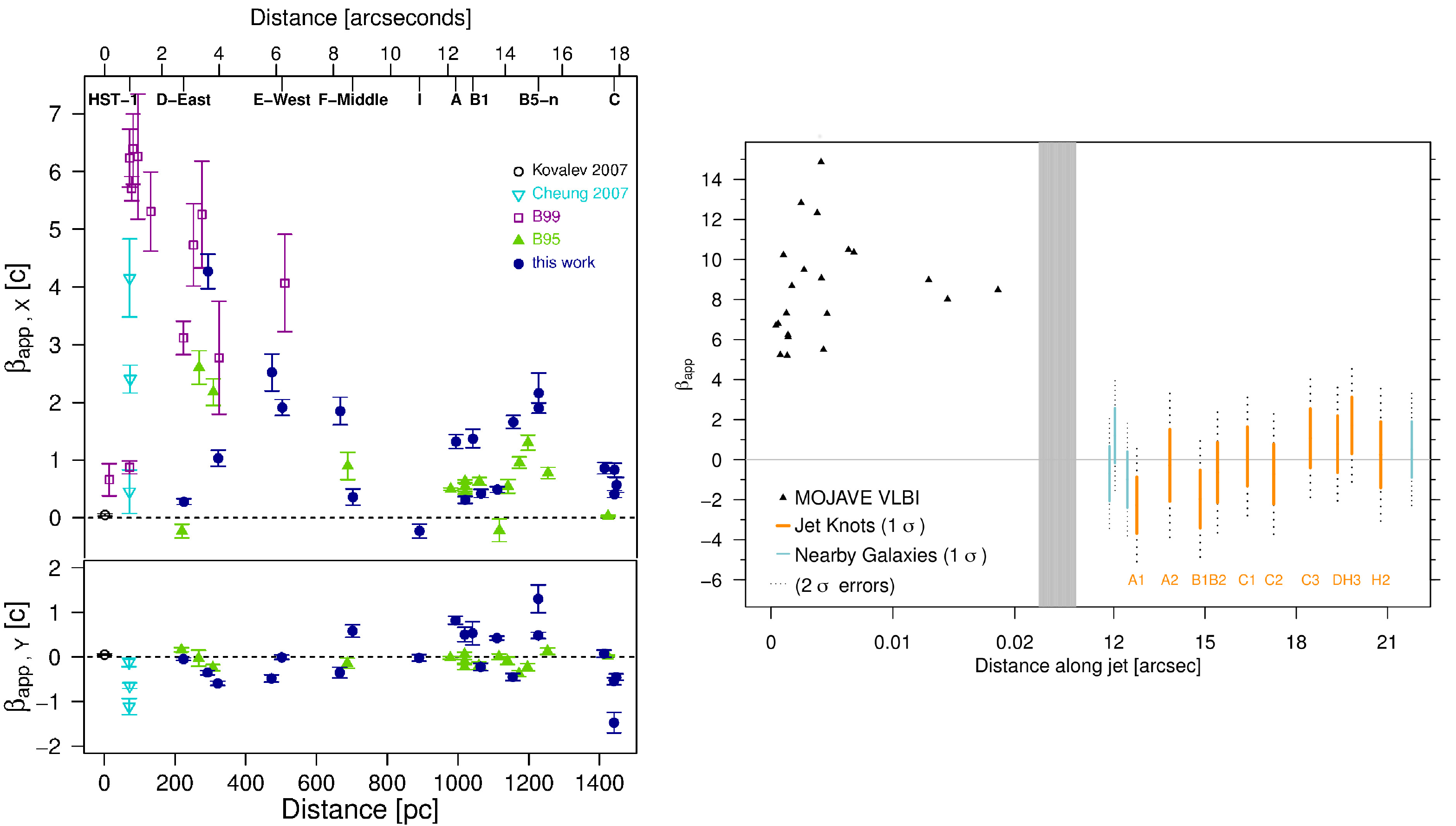}
\caption{Apparent speed versus distance for M87 ({\em left}) and 3C 273 ({\em right}).  For M87, the most recent HST 
	 results are in dark blue.  Note the fast rise in speeds on scales less than 100 parsecs, reaching a peak at HST-1 before gradually falling off. In the case of 3C~264, our results suggest a great similarity to the velocity structure seen in M87. Lower panel shows the significant transverse velocities detected in our study. On the right, we compare the VLBI 
	 scale motions for 3C~273 ({\em left-half}) from the MOJAVE program to the HST proper motions ({\em right-half}) on kpc scales (note the break in the displayed range at the grey vertical bar).  }\label{f3}
\end{figure}

On the left in Figure~\ref{f3}, we show the complete radial and transverse velocity field of M87 as a function of distance along the jet, including  previous work covering similar scales and radio interferometry~results.

\subsection{3C~264}

\begin{figure}[H]
\centering
\includegraphics[width=14cm]{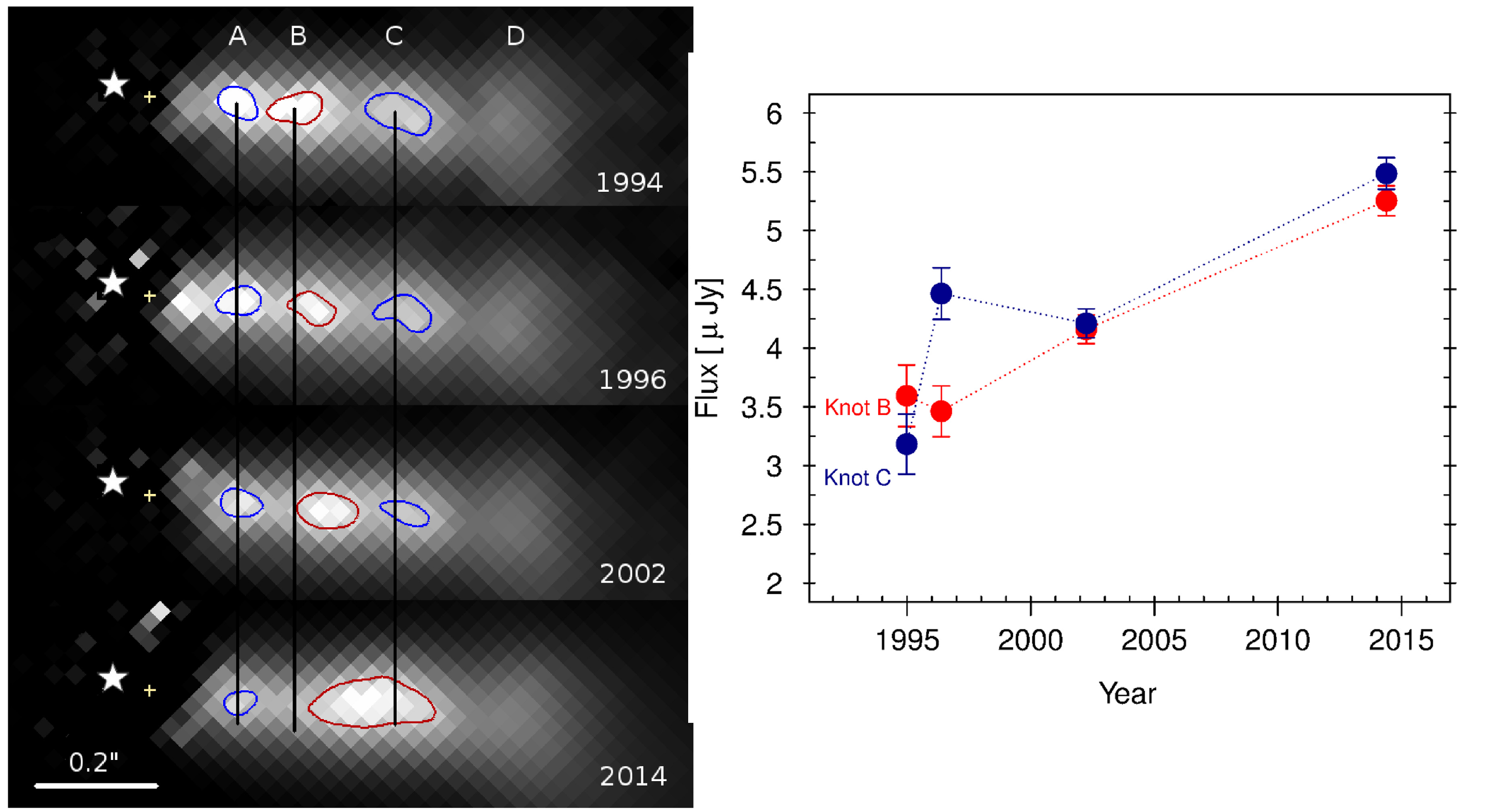}
\caption{Results for 3C~264. Previous radio observations have revealed
that the initially narrowly collimated jet bends by $\sim$10$^\circ$ at
the location marked by the yellow cross~\citep{lara04}, which appears
to align well with the central axis of the jet in our imaging, and
which serves as our reference point for all measured positions. On the left, the fast motion of knot B and subsequent collision knot C and brightening can be clearly seen over the 20 year study.  On the right, we show the increasing flux of the two colliding~knots.  }\label{f4}
\end{figure}

After aligning all images to a common reference frame, the fast proper motion of knot B is clearly
visible, as shown in Figure~\ref{f4}.
We found that knots A and D have a $\beta_{app}$ consistent with zero, while the inner knots B and C have
$\beta_{app}$~=~7.0~$\pm$~0.8$c$ and 1.8~$\pm$~0.5$c$, respectively
(Figure~\ref{f2}). The value for knot B exceeds the fastest speeds measured
in the jet in M87, the only other source for which speeds on kpc
scales have been measured \citep{biretta1999,mey13_m87}. The difference in
speeds between knots B and C puts them on a collision course, an
interaction which has already begun in the final epoch from 2014,
where the knots begin to overlap (Figure~\ref{f2}). This appears to be the first direct observation of colliding knots in an extragalactic jet.

Our data show that in the final 2014 epoch, both knots B and C
brighten at the same time by approximately 40\% over the mean flux
level of the previous three epochs (Figure~\ref{f4}, right). This suggests that
fresh particle acceleration has recently taken place, which we
 attribute to the ongoing collision between these components. Under
equipartition, the cooling length for the optical-emitting electrons
for knots B and C is longer than the distance they travelled in our
observations, consistent with the lack of any decay in flux levels for
these knots over the first three epochs.  This is not the case for
stationary knot A, which can be seen to decay with a timescale of
$\sim$70 years. Knot A appears analogous to knot HST-1 in the M87 jet;
the latter is thought to be a stationary reconfinement shock where the
jet pressure drops below that of the external
environment~\citep{stawarz06}. The event that energized knot A may have
been the passage of fast-moving knot B circa 1971$^{+8}_{-17}$,
comparable to the knot A decay time.

Radio data taken in October 1983 with the VLA suggest that knots B and C were moving faster
in the past and may have decelerated.
In the case of knot B, a quadratic fit to the HST + VLA data suggests
that in late 1983, the knot had a speed of 10.2$c$, slowing by
0.16$c$/year to reach 5.6$c$ at the beginning of 2014.  While a
quadratic fit is very poor for the combined data on knot C, a linear
fit between the 1983 and 1994 epochs similarly suggests a speed of
9.8$c$. It appears that a much more distinct deceleration ``event''
occurred for knot C.

In the internal shock model, the efficiency ($\eta$) of the conversion
of the liberated kinetic energy ($E_{kin}$) into radiation is
generally unknown, with values assumed in the literature in order to
match the source luminosity. For the collision in 3C 264, we can
estimate the energy available for conversion from our observations to be at least 10$^{-3}$, subject to uncertainties in the total cooling time, and the future evolution of the observed flux.

\subsection{3C~273}

In the case of 3C~273, all knots
have speeds consistent with zero with typical 1$\sigma$ errors on the
order of 0.1$-$0.2 mas/year or 1.5$c$, and with 99\% upper limit
values ranging from $1-5$c. Nearby background galaxies, used as a control check,
show that these limits are consistent with stationary objects in the same field (see Figure~3, right).

These results suggest that the knots in the kpc-scale jet, if they are
moving packets of plasma, must~be relatively slow, in
  agreement with previous studies based on jet-to-counterjet ratios in radio-loud populations \citep{arshakian2004,mullin2009}.  Assuming
  that the jet either remains at the same speed or decelerates as you
  move downstream, the $2\sigma$ upper limit speed derived from all
  knots combined of $1c$ suggests that the entire optical jet is at
  most mildly relativistic, with a maximum Lorentz factor of
  $\Gamma<2.9$. However, we~cannot rule out the possibility that the
  knots are standing shock features in the flow, where the bulk plasma
  moves through the features with a higher $\Gamma$.  The best limits
  on the bulk plasma speed thus remain the limits derived from the
  non-detection of the IC/CMB 
   component in gamma-rays by
  \cite{mey15_iccmb}, where $\delta<7.8$ is implied assuming
  equipartition magnetic fields.

Finally, we show that the observed upper limits on the proper motion
of the knots confirms that the a near-equipartition IC/CMB model for
the X-rays of the kpc-scale knots is ruled out. The~equipartition
IC/CMB model requires that the knots are ballistic packets of moving
plasma moving at the bulk speed $\Gamma\approx15-20$ which would imply
proper motions on the order of 10$c$ or 1.12 mas/year which could have
been detected in our study; our upper limits easily rule this out at a
high level of significance ($>$5$\sigma$). Moving away from
equipartition conditions, an IC/CMB model consistent with our
observations requires a jet power on the order of five to several
hundred times the Eddington limit, and is thus energetically disfavored.

In comparison to other recent HST observations of lower-power optical
jets M87 and 3C264, where highly superluminal speeds (6$-$7$c$) have been
observed in the optical kpc-scale jet, our first proper-motion study
of a powerful quasar jet reveals no significant
proper motions.  It remains to be seen whether this is
because the jet has truly decelerated and is only mildly relativistic,
or because the knot features in sources such as M87/3C~264 and 3C~273 represent
very different things: moving packets of plasma in the first instance
and standing shocks in the second.

\vspace{6pt}


\acknowledgments{Eileen T. Meyer acknowledges HST grants AR-11283, GO-12271, AR-12635, GO-13327}

\authorcontributions{Eileen T. Meyer reduced the data and wrote the article.  Jay Anderson, Tony Sohn, and Roeland van der Marel assisted with the geometric correction and astronomy portion of the analysis. All authors contributed to the interpretation and have read and commented on the manuscript.}

\conflictofinterests{The authors declare no conflict of interest.}


\bibliographystyle{mdpi}

\renewcommand\bibname{References}



\end{document}